\documentclass[prl,balancelastpage,twocolumn,footinbib,superscriptaddress,showpacs]{revtex4-1}
\usepackage{color,amsthm,amsmath,amsfonts,graphicx,bm}
\usepackage{bm}
\usepackage{color}
\usepackage{amsfonts}
\usepackage{amssymb}
\usepackage{amsmath}
\usepackage{epsfig}
\usepackage{graphicx}
\usepackage{enumerate}
\usepackage{multirow}
\usepackage{verbatim}
\usepackage{color}
\usepackage{subfigure}

\usepackage[breaklinks=true,colorlinks=true,linkcolor=blue,urlcolor=blue,citecolor=blue]{hyperref}
\begin{document}


\title{Atom-Photon Spin-Exchange Collisions Mediated by Rydberg Dressing}
\author{Fan Yang}
\affiliation{State Key Laboratory of Low Dimensional Quantum Physics, Department of Physics, Tsinghua University, Beijing 100084, China}
\author{Yong-Chun Liu}
\email{ycliu@mail.tsinghua.edu.cn}
\affiliation{State Key Laboratory of Low Dimensional Quantum Physics, Department of Physics, Tsinghua University, Beijing 100084, China}
\author{Li You}
\email{lyou@mail.tsinghua.edu.cn}
\affiliation{State Key Laboratory of Low Dimensional Quantum Physics, Department of Physics, Tsinghua University, Beijing 100084, China}
\affiliation{Frontier Science Center for Quantum Information, Beijing 100084, China}
\affiliation{Beijing Academy of Quantum Information Sciences, Beijing 100193, China}


\begin{abstract}
We show that photons propagating through a Rydberg-dressed atomic ensemble can exchange its spin state with a single atom. Such a spin-exchange collision exhibits both dissipative and coherent features, depending on the interaction strength. For strong interaction, the collision dissipatively drives the system into an entangled dark state of the photon with an atom. In the weak interaction regime, the scattering coherently flips the spin of a single photon in the multi-photon input pulse, demonstrating a generic single-photon subtracting process. An analytic analysis of this process reveals a universal trade-off between efficiency and purity of the extracted photon, which applies to a wide class of single-photon subtractors. We show that such a trade-off can be optimized by adjusting the scattering rate under a novel phase-matching condition.
\end{abstract}

\maketitle

Achieving strong light-atom interaction at the single-particle level represents a long-standing goal in quantum optics \cite{cirac2017quantum}. Realizing this goal will not only enable one to test fundamental physics in quantum electrodynamics (QED) \cite{rempe1987,zhu1990vacuum,thompson1992observation,brune1996quantum,raimond2001manipulating}, but also facilitate meaningful applications of quantum communication \cite{kimble2008quantum,sangouard2011quantum,reiserer2015cavity}, simulation \cite{gonzalez2015subwavelength,douglas2015quantum,hung2016quantum,chang2018colloquium}, and metrology \cite{haas2014entangled,mcconnell2015entanglement,masson2017cavity}. As a promising approach, interfacing photons with Rydberg atoms \cite{saffman2010quantum} via electromagnetically induced transparency (EIT) \cite{fleischhauer2005electromagnetically} has attracted much attention in recent years \cite{pritchard2010cooperative,gorshkov2011photon,peyronel2012quantum,baur2014single,gorniaczyk2014single,tresp2016single,parigi2012observation,tiarks2016optical,tiarks2019photon,firstenberg2016nonlinear,thompson2017symmetry,murray2017coherent,tiarks2019photon,khazali2019polariton}.
To date, a host of interaction processes have been established with this approach, e.g., a single atomic excitation can block the transmission of a single photon \cite{pritchard2010cooperative,gorshkov2011photon,peyronel2012quantum,baur2014single,gorniaczyk2014single}, imprint a global phase onto a single photon \cite{parigi2012observation,tiarks2016optical,tiarks2019photon}, reflect a single photon \cite{murray2017coherent}, or exchange its position with a single photon \cite{thompson2017symmetry,khazali2019polariton}.

In this Letter, we establish a different type of atom-photon interaction in the Rydberg EIT system, with which a single photon can exchange its spin state with a single atom. It is achieved by coupling photons to an atomic ensemble that interacts with a single control atom via Rydberg dressing \cite{glaetzle2017quantum,yang2019quantum}. We show that under suitable conditions, the scattering dynamics can be tuned from dissipative to coherent. In the dissipative regime, the system evolves robustly into an entangled dark state of a photon and the control atom. For coherent scattering, the dynamics maps to a model of generic single-photon subtraction, whose solution reveals a universal trade-off between efficiency and purity of the subtracted single photon, and yields a phase-matching condition for optimizing its performance.

\begin{figure}[b]
\centering
\includegraphics[width=\linewidth]{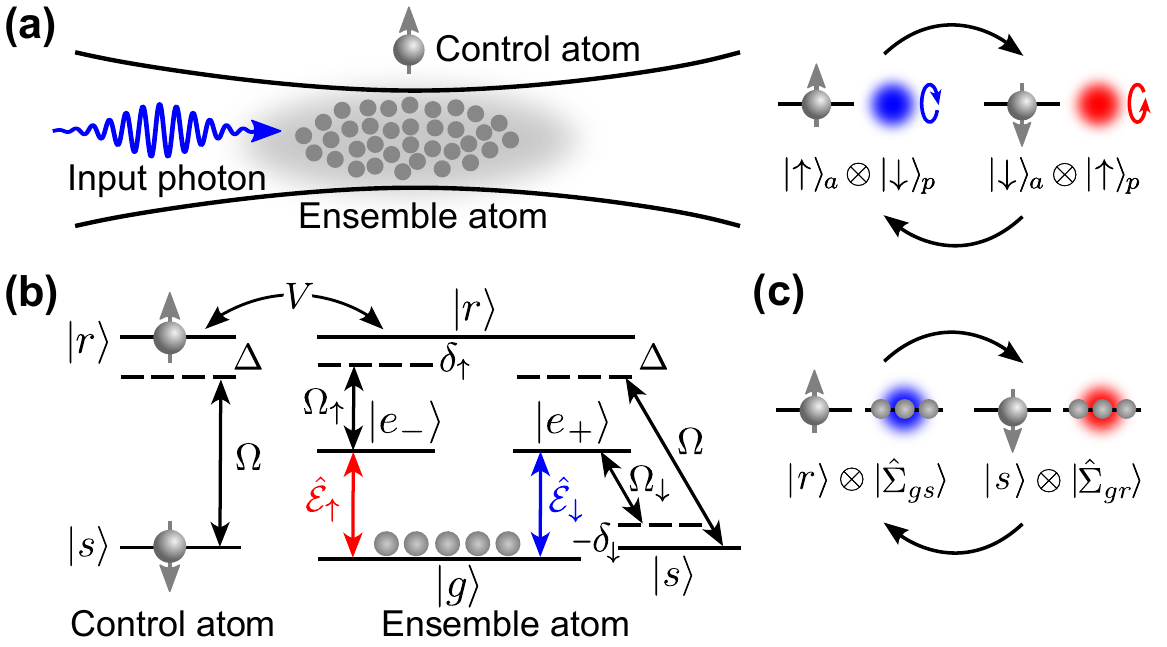}
\caption{(a) Schematic for the spin-exchange collision between input photons and the control atom. (b) Level structure for the control atom and the ensemble atom. For $^{87}$Rb atom, we can choose $|g\rangle = |5S_{1/2}, F =1, m_F = 0\rangle$, $|e_\pm\rangle = |5P_{3/2}, F =2, m_F =\pm1\rangle$, $|s\rangle = |5S_{1/2}, F =2, m_F = 0\rangle$, and $|r\rangle = |nS_{1/2}, J =1/2, m_J = -1/2\rangle$. The coupling $\Omega$ between $|s\rangle$ and $|r\rangle$ can be constructed using a two-photon process with an intermediate state $|5P_{1/2},F =1,m_F =-1\rangle$. The two-photon detunings are $\delta_\downarrow=\Omega^2/\Delta$ and $\delta_\uparrow=-\delta_\downarrow$. (c) Schematic of the spin-exchange between the control atom and a spin-wave excitation in the ensemble.}
\label{fig:fig1}
\end{figure}

The system we study is illustrated in Fig.~\ref{fig:fig1}(a), where the input photon carries photonic spin (polarization) and can exchange its state with the pseudo-spin (internal state) of the control atom. This atom-photon spin-exchange interaction is mediated by an atomic ensemble, which strongly interacts with both the photon and the control atom \cite{saffman2005entangling,petrosyan2018deterministic,grankin2018free}. The level structure shown in Fig.~\ref{fig:fig1}(b) helps to realize such an interaction. A photon propagates in an atomic ensemble via two distinct EIT processes \cite{ruseckas2018nonlinear,yang2019manipulating}: the left circularly polarized (pseudo-spin up) photonic field $\hat{\mathcal{E}}_\uparrow(\mathbf{r})$ forms a Rydberg EIT involving the ground state $|g\rangle$, the intermediate state $|e_-\rangle$, and the Rydberg state $|r\rangle$; while the right circularly polarized (pseudo-spin down) photonic field $\hat{\mathcal{E}}_\downarrow(\mathbf{r})$ participates in a $\Lambda$-type EIT formed by $|g\rangle$, $|e_+\rangle$, and another ground state $|s\rangle$. In addition, state $|s\rangle$ is dressed
to Rydberg state $|r\rangle$ for both the control atom and ensemble atoms.

It is shown in Ref.~\cite{yang2019quantum} that the above dressing scheme induces an effective spin-exchange interaction $\hat{V}_\mathrm{ex}$ between atoms in $|s\rangle$ and $|r\rangle$. At low photon density, interactions between ensemble atoms are negligible, such that $\hat{V}_\mathrm{ex}=\sum_iU(\mathbf{r}_i)\hat{\sigma}_{rs}\hat{\sigma}_{sr}^i+\mathrm{H.c.}$ just describes the spin-exchange between the control atom ($\hat{\sigma}_{\mu\nu}=|\mu\rangle\langle\nu|$) and each $i$-th atom ($\hat{\sigma}_{\mu\nu}^i$) in the ensemble. Since most ensemble atoms are in the ground state $|g\rangle$, $\hat{V}_\mathrm{ex}$ actually describes the spin-exchange between the control atom and a collective excitation (spin-wave) in the atomic ensemble [Fig.~\ref{fig:fig1}(c)], i.e., $\hat{V}_\mathrm{ex}=
\int d\mathbf{r}U(\mathbf{r})\hat{\sigma}_{rs}\hat{\Sigma}_{gs}^\dagger(\mathbf{r})
\hat{\Sigma}_{gr}(\mathbf{r})+\mathrm{H.c.}$, where $\hat{\Sigma}_{g\mu}(\mathbf{r})$ denotes the spin-wave field operator for the collective excitation in state $|\mu\rangle$ \cite{supply}. With the above EIT configuration, the spin-wave field $\hat{\Sigma}_{gr}(\mathbf{r})$ is coupled to the photonic field $\hat{\mathcal{E}}_\uparrow(\mathbf{r})$ to form a dark state polariton (DSP), while $\hat{\Sigma}_{gs}(\mathbf{r})$ is coupled to $\hat{\mathcal{E}}_\downarrow(\mathbf{r})$ to form another DSP. In this way, $\hat{V}_\mathrm{ex}$ maps to the exchange interaction between the control atom and the photonic field.

The exchange interaction takes the form $U(\mathbf{r})=U_0/[1+(|\mathbf{r}|/R_c)^6]$, where the strength $U_0=\Omega^2/\Delta$ is determined by the Rabi frequency $\Omega$ and the detuning $\Delta$ of the dressing field ($\Omega\ll\Delta$),  and the effective range is $R_c=(C_6/\Delta)^{1/6}$ with $C_6$ the van der Waals (vdW) interaction coefficient between atoms in state $|r\rangle$ \cite{yang2019quantum}. It does not need one to tune near a F\"{o}ster resonance and can be conveniently controlled by the dressing field. Furthermore, the dressing scheme adopted here suppresses the unwanted direct interaction ($\sim\Omega^4/\Delta^3$) between input photons in mode $\hat{\mathcal{E}}_\downarrow$. These desirable features as well as other details are compared to the off-diagonal vdW interaction scheme in the Supplemental Material \cite{supply}.

{\it Single-photon scattering}.---First, we consider the interaction between the control atom and a single photon propagating along $z$-direction. Neglecting the decoherence of the Rydberg state, the input/output state in the one-dimensional (1D) case can be expressed as
\begin{align}
|\psi(t)\rangle = &\int{dz}E_{\downarrow\uparrow}(z,t)\hat{\mathcal{E}}_{\downarrow}^\dagger(z)|0\rangle\otimes |\hspace{-2.5pt}\uparrow\rangle_{a}\nonumber\\
&+ \int{dz}E_{\uparrow\downarrow}(z,t)\hat{\mathcal{E}}_{\uparrow}^\dagger(z)|0\rangle\otimes |\hspace{-2.5pt}\downarrow\rangle_{a},
\label{eq:eq1}
\end{align}
where $|0\rangle$ denotes the vacuum state for photons, and $|$$\uparrow\rangle_{a}=|r\rangle$, $|$$\downarrow\rangle_{a}=|s\rangle$ represent two internal states of the control atom. The spatiotemporal feature of the photon is described by the wavefunction $E_{\mu\nu}(z,t)=\langle\nu|_{a}\langle0|\hat{\mathcal{E}}_{\mu}(z)|\psi(t)\rangle$. The output state of the system is determined by the dynamics inside the atomic ensemble $z\in[0,L]$, where the spin-wave field needs to be taken into consideration. Let $P_{\downarrow\uparrow}$, $S_{\downarrow\uparrow}$, $P_{\uparrow\downarrow}$, and $S_{\uparrow\downarrow}$ describe the collective excitation in state $|e_+\rangle$, $|s\rangle$, $|e_-\rangle$, and $|r\rangle$, respectively \cite{definition}. Then the evolution of the wavefunction $\psi(z,t)=(E_{\downarrow\uparrow},P_{\downarrow\uparrow},S_{\downarrow\uparrow},E_{\uparrow\downarrow},P_{\uparrow\downarrow},S_{\uparrow\downarrow})^T$ is governed by $i\partial_t\psi=\mathcal{H}\psi$ \cite{supply} with
\begin{equation}
\mathcal{H}=\begin{bmatrix}
-ic\partial_z&g_p&0&0&0&0\\
g_p&-i\gamma&\Omega_\downarrow&0&0&0\\
0&\Omega_\downarrow&U(z)&0&0&U(z)\\
0&0&0& -ic\partial_z&g_p&0\\
0&0&0&g_p&-i\gamma&\Omega_\uparrow\\
0&0&U(z)&0&\Omega_\uparrow&U(z)
\end{bmatrix},
\label{eq:eq2}
\end{equation}
where $g_p$ and $2\gamma$ are the collective atom-photon coupling constant and the linewidth of the $|g\rangle-|e_\pm\rangle$ transition, respectively, $\Omega_\uparrow$ ($\Omega_\downarrow$) denotes the control field for the Rydberg ($\Lambda$-type) EIT, and $U(z)=U_0/[1+(\sqrt{z^2+r_\perp^2}/R_c)^6]$ is the potential. In the frequency ($\omega$) domain, we have
\begin{align}
i\partial_z\begin{bmatrix}
{{E}}_{\downarrow\uparrow}\\
{{E}}_{\uparrow\downarrow}
\end{bmatrix}=\begin{bmatrix}
\chi_\downarrow(z,\omega)&\kappa(z,\omega)\\
\kappa(z,\omega)&\chi_\uparrow(z,\omega)
\end{bmatrix}
\begin{bmatrix}
{{E}}_{\downarrow\uparrow}\\
{{E}}_{\uparrow\downarrow}
\end{bmatrix},
\label{eq:eq3}
\end{align}
where the susceptibilities $\chi_{\downarrow}$ and $\chi_{\uparrow}$ come from the dressing induced diagonal interaction, while $\kappa$ describes the spin-exchange coupling between states $|$$\downarrow\rangle_{p}\otimes|$$\uparrow\rangle_{a}$ and $|$$\uparrow\rangle_{p}\otimes|$$\downarrow\rangle_{a}$. For the input state $|$$\downarrow\rangle_{p}\otimes|$$\uparrow\rangle_{a}$, the solution to Eq.~(\ref{eq:eq3}) can be written as ${E}_{\downarrow\uparrow}(L,\omega)=T(\omega){E}_{\downarrow\uparrow}(0,\omega)$, and ${E}_{\uparrow\downarrow}(L,\omega)=R(\omega){E}_{\downarrow\uparrow}(0,\omega)$. At steady state ($\omega=0$), we find $\chi_{\mu}(z,0)=\mathcal{V}(z)/v_\mu$, and $\kappa(z,0)=\mathcal{V}(z)/\sqrt{v_\uparrow v_\downarrow}$, where $v_\mu=c\Omega_\mu^2/g_p^2$ ($\mu=\uparrow,\downarrow$) is the photon group velocity in the slow-light regime, and
\begin{equation}
\mathcal{V}(z) = \frac{U(z)}{1+i\gamma U(z)(\Omega_\downarrow^2+\Omega_\uparrow^2)/\Omega_\downarrow^2\Omega_\uparrow^2}
\end{equation}
is the effective potential. In this case, the scattering coefficients $T(0)=(\Omega_\uparrow^2e^{-i2\phi}+\Omega_\downarrow^2)/(\Omega_\downarrow^2+\Omega_\uparrow^2)$ and $R(0)=\Omega_\uparrow\Omega_\downarrow (e^{-i2\phi}-1)/(\Omega_\downarrow^2+\Omega_\uparrow^2)$ are determined by the interaction induced phase factor $\phi=(v_\uparrow+v_\downarrow)\int_0^Ldz\mathcal{V}(z)/2v_\uparrow v_\downarrow$. For $L>4R_c$ and $r_\perp<R_c$, the complex phase factor is simply given by $\phi\approx(2\pi/3)\xi[1-i(5/3)\xi]\times\mathrm{OD}_c$, where $\xi=U_0/\gamma_\mathrm{EIT}$ measures the interaction strength in units of the effective EIT linewidth $\gamma_\mathrm{EIT}=2\Omega_\uparrow^2\Omega_\downarrow^2/(\Omega_\downarrow^2+\Omega_\uparrow^2)\gamma$, and $\mathrm{OD}_c=g_p^2R_c/\gamma c$ denotes the effective optical depth.

\begin{figure}
\centering
\includegraphics[width=\linewidth]{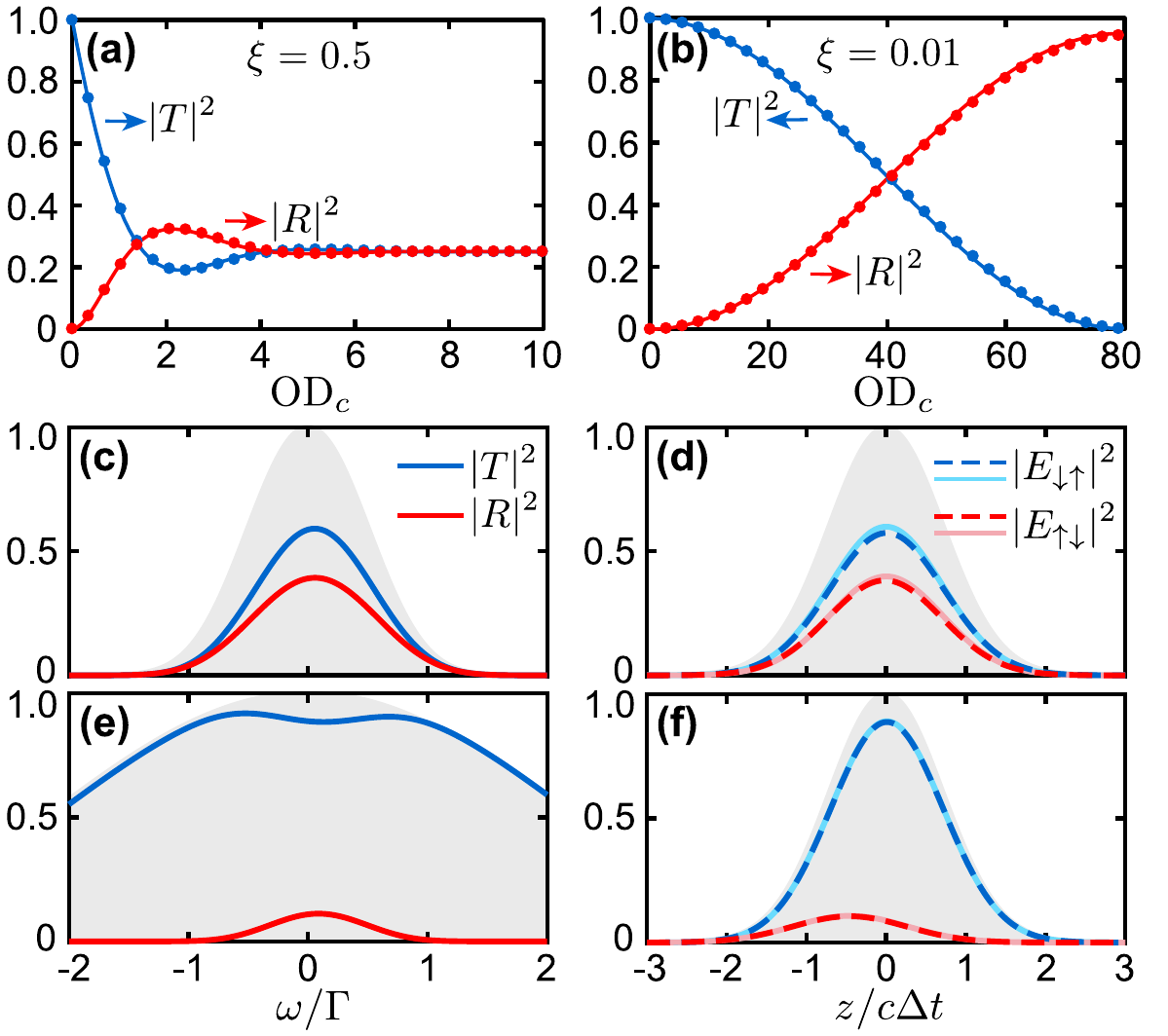}
\caption{(a) and (b) Scattering coefficients versus $\mathrm{OD}_c$ in the dissipative ($\xi=0.5$) and the coherent ($\xi=0.01$) regime. The dots and the solid lines correspond to the results for a Gaussian beam with a waist $w$ and the $1$D model, respectively. We take $\Omega_{\uparrow,\downarrow}/2\pi=3~\mathrm{MHz}$ and $R_c=9~\mu\mathrm{m}$ in (a), $\Omega_{\uparrow,\downarrow}/2\pi=8~\mathrm{MHz}$ and $R_c=12~\mu\mathrm{m}$ in (b), $\gamma/2\pi=3~\mathrm{MHz}$, $r_\perp =2w=4~\mu\mathrm{m}$, $L=4R_c$, and $\Delta=10\Omega$.\hspace{-1pt} (c)-(f) Spectra of the scattering coefficients in units of the EIT bandwidth $\Gamma=\Omega_\uparrow^2/\gamma\sqrt{\mathrm{OD}}$ [$\mathrm{OD}=(L/R_c)\mathrm{OD}_c$] and evolution of the wavefunctions for a Gaussian input pulse with a duration $\Delta t=10/\Gamma$. The shaded areas denote the non-interacting transmission. The dashed and the solid lines in (d) and (f) are based on Eq.~(\ref{eq:eq3}) and Eq.~(\ref{eq:eq5}), respectively. The parameters are the same as in (b) ($\mathrm{OD}_c=35$) except that we introduce a group velocity mismatch in (e) and (f) by taking $\Omega_\downarrow/2\pi=16~\mathrm{MHz}$.}
\label{fig:fig2}
\end{figure}

When the interaction strength $U_0$ is comparable to the EIT linewidth $\Omega_\uparrow^2/\gamma$ or $\Omega_\downarrow^2/\gamma$, the ratio $\xi$ is large. Consequently, both $\mathcal{V}(z)$ and $\phi$ have a large imaginary part. In this dissipative interacting regime, as $\mathrm{OD}_c$ increases, the photon loss probability rapidly grows. However, Eq.~(\ref{eq:eq3}) possesses an eigen state free from dissipation, such that as $\mathrm{OD}_c$ increases further, the loss rate saturates, and the system eventually evolves into this dark state $(\Omega_\downarrow|$$\downarrow\rangle_{p}|$$\uparrow\rangle_{a}-
\Omega_\uparrow|$$\uparrow\rangle_{p}|$$\downarrow\rangle_{a})/(\Omega_\downarrow^2+\Omega_\uparrow^2)^{1/2}$ with a probability $\Omega_\downarrow^2/(\Omega_\downarrow^2+\Omega_\uparrow^2)$ [Fig.~\ref{fig:fig2}(a)]. Thus, such a dissipative spin-exchange collision can be used for robust generation of atom-photon entanglement. If the interaction strength is much smaller than the EIT linewidth, i.e., $\xi\ll1$, the effective potential $\mathcal{V}(z)\approx U(z)$ is essentially real and the imaginary part of $\phi$ is largely suppressed. In this case, as $\mathrm{OD}_c$ increases, the system undergoes a coherent oscillation between $|$$\downarrow\rangle_{p}\otimes|$$\uparrow\rangle_{a}$ and $|$$\uparrow\rangle_{p}\otimes|$$\downarrow\rangle_{a}$ [Fig.~\ref{fig:fig2}(b)]. We calculate the scattering coefficients for a finite beam width $w<R_c$ \cite{khazali2019polariton} [see Figs.~\ref{fig:fig2}(a) and \ref{fig:fig2}(b)], and find nice agreement with the results predicted by the 1D model.


We now focus on the coherent scattering process. When dissipations are negligible as in this case, the propagation of photons inside the atomic ensemble can be described by DSP fields \cite{fleischhauer2000dark} $\hat{\Psi}_\uparrow(z)=\cos\theta_\uparrow\hat{\mathcal{E}}_\uparrow(z)-\sin\theta_\uparrow\hat{\Sigma}_{gr}(z)$ and $\hat{\Psi}_\downarrow(z)=\cos\theta_\downarrow\hat{\mathcal{E}}_\downarrow(z)-\sin\theta_\downarrow\hat{\Sigma}_{gs}(z)$ with $\tan\theta_\mu=g_p/\Omega_\mu$. As verified by Figs.~\ref{fig:fig2}(c)-\ref{fig:fig2}(f), for frequency components well within the EIT bandwidth, the dynamics inside the medium can be described by the following Hamiltonian
\begin{align}
\hat{H}=&-iv_{\downarrow}\int dz\hat{\Psi}^\dagger_{\downarrow}(z)\partial_z\hat{\Psi}_{\downarrow}(z)-iv_{\uparrow}\int dz\hat{\Psi}^\dagger_{\uparrow}(z)\partial_z\hat{\Psi}_{\uparrow}(z)\nonumber\\
&+\int{dz}U(z)\left[\hat{\sigma}_{\uparrow\uparrow}\hat{\Psi}_{\downarrow}^\dagger(z)\hat{\Psi}_{\downarrow}(z) +\hat{\sigma}_{\downarrow\downarrow}\hat{\Psi}_{\uparrow}^\dagger(z)\hat{\Psi}_{\uparrow}(z)\right]\nonumber\\
&+\int{dz}U(z)\left[\hat{\sigma}_{\uparrow\downarrow}\hat{\Psi}_{\downarrow}^\dagger(z)\hat{\Psi}_{\uparrow}(z)+\mathrm{H.c.}\right],
\label{eq:eq5}
\end{align}
whose first line denotes the photon kinetic energy, and the second (third) line represents the density (spin-exchange) interaction between the photon and the atom.

The single-photon scattering elucidated above can be used as a building block in quantum networks. At small $\mathrm{OD}_c$, the induced atom-photon entanglement can be further purified to establish quality entanglement between distant atoms \cite{supply}. Unlike the DLCZ protocol \cite{duan2001long}, the entanglement we discuss here refers to polarization bases instead of Fock space, so that photon-number resolved detectors are not required, and the system is insensitive to interferometric instabilities \cite{chen2007fault,sangouard2011quantum}. At large $\mathrm{OD}_c$ that gives $\phi=\pi/2$, the spin-exchange collision leads to a direct mapping between atomic and photonic states if $\Omega_\uparrow=\Omega_\downarrow$, i.e., $|$$\downarrow\rangle_p\otimes(\alpha|$$\downarrow\rangle+\beta|$$\uparrow\rangle)_a\leftrightarrow(\alpha|$$\downarrow\rangle
-\beta|$$\uparrow\rangle)_p\otimes|$$\downarrow\rangle_a$, which facilitates quantum state transfer in a network.

{\it Multi-photon scattering}.---Next, we consider coherent spin-exchange collisions [governed by Eq.~(\ref{eq:eq5})] between the control atom and an input pulse containing $n$ identical photons. Here, we focus on the limit of a long input pulse with a duration $\Delta t\gg nR_c/v_\mu$. In this low-photon-density regime, photons rarely interact with the control atom at the same time, which allows us to obtain an analytical form for the output state based on single-photon scattering coefficients, without numerically solving the multi-photon Schr\"{o}dinger equation based on Eq.~(\ref{eq:eq5}).

Assuming the $n$ incoming photons are in the spin-down state with a real temporal wavefunction $h(t)$ normalized as $\int dth^2(t)=1$ and the control atom is initially spin-up, the input state of the system is given by (taking $c=1$)
\begin{align}
|\psi_\mathrm{in}(t)\rangle&=\frac{1}{\sqrt{n!}}\left[\int_{-\infty}^{\infty}{dz}h(t-z)\hat{\mathcal{E}}_{\downarrow}^\dagger(z)\right]^n|0\rangle|\hspace{-2.5pt}\uparrow\rangle_{a}\nonumber\\
&=\sqrt{n!}\int_{t_n>\cdots>t_1}\left[\prod_{i=1}^n{dt_i}h(t_i)\hat{\mathcal{E}}_{\downarrow}^\dagger(t-t_i)\right]|0\rangle|\hspace{-2.5pt}\uparrow\rangle_{a},\nonumber
\end{align}
where time ordering for the input photons is introduced \cite{gorshkov2013dissipative}. For coherent spin-exchange collisions governed by Eq.~(\ref{eq:eq5}), the total magnetization $\hat{\sigma}_{\uparrow\uparrow}+\int dz \hat{\Psi}^\dagger_\uparrow(z)\hat{\Psi}_\uparrow(z)=1$ is conserved, which implies that at most one of the photons can be scattered to flip its spin state. At low photon density, photons interact with the atom one after the other, i.e., if a photon propagates through the medium without exchanging its state with the atom, the next photon still has a probability to do so; but once the exchange occurs, the remaining photons will keep their spin states. In this way, the output state is given by
\begin{equation}
|\psi(t)\rangle=T^{n}|\psi_\mathrm{in}(t-\tau)\rangle+\sqrt{n!}\sum_{m=1}^{n}RT^{m-1}|\psi_{m}(t)\rangle,
\label{eq:eq6}
\end{equation}
where $|\psi_\mathrm{in}(t-\tau)\rangle$ corresponds to the situation in which no spin-exchange occurs, while $|\psi_m(t)\rangle$ denotes the event that the spin-exchange is between the control atom and the $m$-th photon in the pulse, given by
\begin{align}
|\psi_{m}(t)\rangle&=\int_{t_n>\cdots>t_{m+1}}\left[\prod_{i=m+1}^n{dt_i}h(t_i)\hat{\mathcal{E}}_{\downarrow}^\dagger(t-\tau^\prime-t_i)\right]\nonumber\\
&\times\int_{-\infty}^{t_{m+1}}dt_mh(t_m)\hat{\mathcal{E}}_{\uparrow}^\dagger(t-\tau-t_m)\nonumber\\
&\times\int_{t_{m}>\cdots>t_1}\left[\prod_{i=1}^{m-1}{dt_i}h(t_i)\hat{\mathcal{E}}_{\downarrow}^\dagger(t-\tau^\prime-t_i)\right]|0\rangle|\hspace{-2.5pt}\downarrow\rangle_{a},\nonumber
\end{align}
with $\tau$ and $\tau^\prime$ the EIT-induced delay time for spin-up and spin-down photons in the atomic ensemble, respectively. In fact, the spin-exchange collision here can be viewed as a heralded single-photon subtractor: a single-photon is subtracted from mode $\hat{\mathcal{E}}_{\downarrow}$ and added to mode $\hat{\mathcal{E}}_{\uparrow}$, conditioned on the spin-flip of the control atom. In contrast to previous schemes \cite{honer2011artificial,murray2018photon}, the single-photon is coherently extracted from the multi-photon pulse here, so it simultaneously behaves as a single-photon source \cite{rosenblum2011photon,rosenblum2016extraction}.

Since the extracted single-photon and the remaining $n-1$ spin-down photons together constitute a pure state, the performance of such a single-photon subtractor can be measured by either part of the system. Tracing out $n-1$ photons in mode $\hat{\mathcal{E}}_{\downarrow}$, the reduced density matrix operator for the spin-up single-photon is $\hat{\rho}=\int dxdy\rho(x,y)\hat{\mathcal{E}}_{\uparrow}^\dagger(x)|0\rangle\langle 0|\hat{\mathcal{E}}_{\uparrow}(y)$, with the density matrix element $\rho(x,y)=\tilde{\rho}(t-\tau-x,t-\tau-y)$ and \cite{note}
\begin{widetext}
\begin{equation}
\tilde{\rho}(x,y)=n|R|^2h(x)h(y)\left[|T|^2\int_{-\infty}^{\min(x,y)}dzh^2(z)+T\int_{\min(x,y)}^{\max(x,y)}dzh^2(z)+\int_{\max(x,y)}^{+\infty}dzh^2(z)\right]^{n-1}.
\label{eq:eq7}
\end{equation}
\end{widetext}

\begin{figure}[b]
\centering
\includegraphics[width=\linewidth]{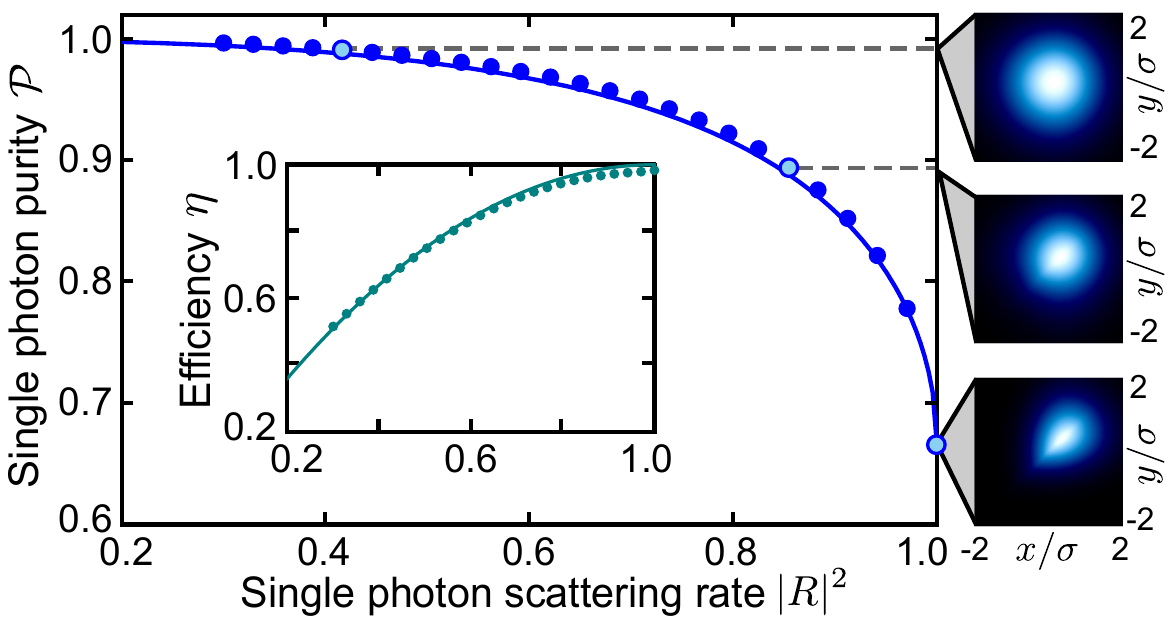}
\caption{Purity and efficiency of the extracted photon. The dots and the solid lines represent numerical ($v_\downarrow\Delta t=20R_c$) and analytical results, respectively. The right figures show the normalized density matrix $|\rho(x,y)|/\eta$ (brighter colors indicate larger values). To assure $\theta=0$, we set $\phi=\pi/2$ and $\Omega_\downarrow>\Omega_\uparrow$.}
\label{fig:fig3}
\end{figure}

The efficiency for scattering a single-photon to spin-up state is found to be $\eta=\mathrm{tr}[\hat{\rho}]=1-|T|^{2n}$, and the purity of this extracted single-photon is given by $\mathcal{P}={\mathrm{tr}[\hat{\rho}^2]}/{\mathrm{tr}[\hat{\rho}]^2}$, which has an analytical expression  
\begin{equation}
\mathcal{P}=\frac{n(1+T)(1-T^{2n-1})}{(2n-1)(1-T^{2n})},
\label{eq:eq8}
\end{equation}
if $T=\sqrt{1-|R|^2}e^{i\theta}$ is real (i.e., $\theta=0,\pi$). For $\theta=0$, this result proves the fundamental trade-off between efficiency and purity of the single-photon subtraction observed in Ref.~\cite{tresp2016single}: while the decrease of the single-photon exchange rate $|R|^2$ reduces the efficiency $\eta$, it yields a larger single-photon purity $\mathcal{P}$. The physical origin of this trade-off comes from entanglement between the subtracted single-photon and the remaining $n-1$ photons. For a perfect exchange $|R|=1$, only $|\psi_1\rangle$ survives in Eq.~(\ref{eq:eq6}), so the timing for the first photon in mode $\hat{\mathcal{E}}_\downarrow$ carries correlated information about the photon in mode $\hat{\mathcal{E}}_\uparrow$. This entanglement results in an impure spin-up photon with $\mathcal{P}=n/(2n-1)$, exactly the case discussed in Ref.~\cite{gorshkov2013dissipative}. In contrast, for $|R|\ll1$ ($T\approx1$), each $|\psi_m\rangle$ in Eq.~(\ref{eq:eq6}) is almost equally weighted, so the timing of the spin-up photon is uncorrelated with the timings of the $n-1$ spin-down photons, i.e., they are not entangled. Therefore, the subtracted photon is almost pure with $\tilde{\rho}(x,y)\sim h(x)h(y)$ and $\mathcal{P}\approx1$. To verify the above analysis, we perform numerical simulations for $n=2$ based on Eq.~(\ref{eq:eq5}). As shown in Fig.~\ref{fig:fig3}, the existence of this trade-off is largely confirmed and good agreement with analytical predictions is observed.

\begin{figure}[b]
\centering
\includegraphics[width=\linewidth]{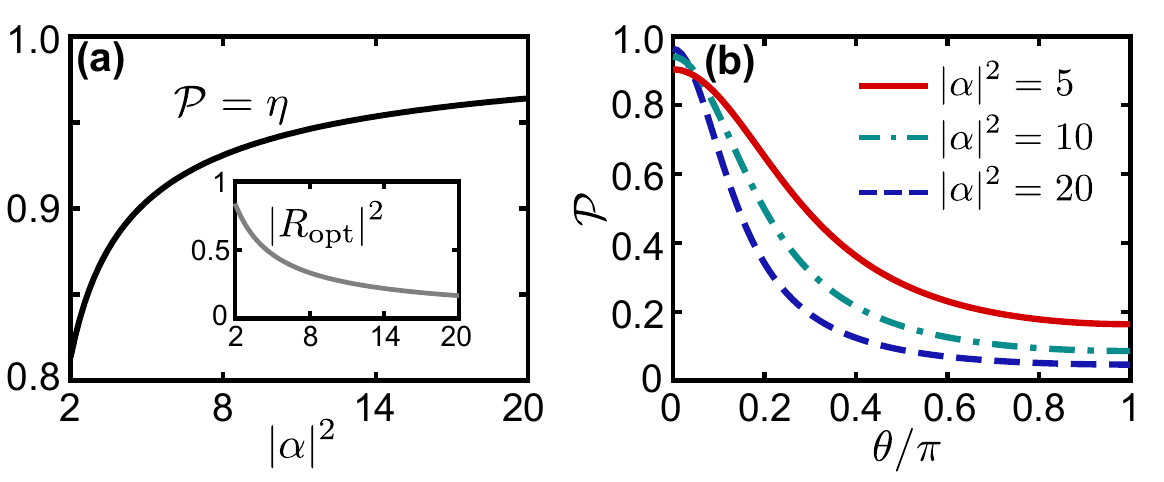}
\caption{(a) Optimized purity and efficiency for the scattering of a coherent input with mean photon number $|\alpha|^2$ ($\theta=0$). The inset shows the optimal scattering rate $|R_\mathrm{opt}|^2$. (b) Purity $\mathcal{P}$ as a function of the phase $\theta$ of the scattering coefficient $T$ at $|R_\mathrm{opt}|^2$ for the indicated value of $|\alpha|^2$.}
\label{fig:fig4}
\end{figure}

We note the above discussed trade-off is universal for a wide class of single photon subtractors in the literature \cite{honer2011artificial,rosenblum2011photon,rosenblum2016extraction,tresp2016single,murray2018photon}, where the arrival order of incoming identical photons is crucial to the output state. Although it prevents the implementation of a perfect single-photon subtraction with $\eta=\mathcal{P}=1$ for arbitrary incoming states, it remains possible to achieve high efficiency and purity simultaneously for a large input photon number. To demonstrate this, we consider the scattering of a coherent input state $e^{-|\alpha|^2/2}\sum_n(\alpha^n/\sqrt{n!})|n\rangle$ with an optimal scattering rate $|R_\mathrm{opt}|^2$ that gives $\eta=\mathcal{P}$. As shown in Fig.~\ref{fig:fig4}(a), both purity and efficiency approach unity as the mean photon number $|\alpha|^2$ increases.

Finally, we emphasize that to achieve the optimal purity, the phase of $T=|T|e^{i\theta}$ needs to be zero, i.e., photons remaining in mode $\hat{\mathcal{E}}_{\downarrow}$ should acquire the same phase irrespective of whether the spin-exchange happens or not. The monotonic decrease of purity [$\mathcal{P}\approx(1-|T|^2)/2(1-|T|\cos\theta)$ for $|\alpha|^2\gg1$] with the phase mismatch $\theta$ [Fig.~\ref{fig:fig4}(b)] can be understood as follows: the phase $(m-1)\theta$ imprinted on $|\psi_m\rangle$ in Eq.~(\ref{eq:eq6}) causes the phase distribution of the spin-up photon strongly correlated with the timing of the remaining photons. In the limit of $|T|\approx1$ and $\theta=\pi$, the purity $\mathcal{P}\approx1/(2n-1)$ is even worse than a perfect exchange, although the probability distribution $\tilde{\rho}(x,x)\sim h^2(x)$ remains unaltered. Such a phase-matching condition highlights the coherent feature of the single-photon subtraction, which cannot be captured by the Monte Carlo simulation used in Ref.~\cite{tresp2016single}.

In conclusion, we present a scheme to engineer spin-exchange interactions between photons and a single atom, and discuss the scattering dynamics for a single-photon as well as a multi-photon input. Further studies can use some recently developed techniques \cite{manzoni2017simulating,zeuthen2017correlated,bienias2018photon,kiilerich2019input} to address the interesting multi-photon scattering problem beyond the low-photon-density regime, where collective effects will come into play. The system can also be used to perform quantum logic operations, such as single-photon optical switching \cite{shomroni2014all}. Besides facilitating quantum information processing, the spin-exchange collision discussed here opens a new avenue for the study of strong light-atom interactions.

\begin{acknowledgments}
We acknowledge valuable discussions with Alexey Gorshkov, Ron Belyansky, and Lin Li. This work is supported by the National Key R$\&$D Program of China (Grant No.~2018YFA0306504) and the National Natural Science Foundation of China (NSFC) (Grants No.~11654001, No.~U1930201, No.~91736106, No.~11674390, and No.~91836302). L.Y. also acknowledges support from BAQIS Research Program (Grant No.~Y18G24).
\end{acknowledgments}

\bibliography{main_text}

\begin{thebibliography}{54}
\bibitem [{def()}]{definition}%
  \BibitemOpen
  \href@noop {} {}\bibinfo {note} {The explicit definitions of these excitation
  wavefunctions are $P_{\downarrow\uparrow}(z,t)=\langle\uparrow$$|_a\langle
  0|\hat{\Sigma}_{ge_+}(z)|\psi(t)\rangle$,
  $S_{\downarrow\uparrow}(z,t)=\langle\uparrow$$|_a\langle
  0|\hat{\Sigma}_{gs}(z)|\psi(t)\rangle$,
  $P_{\uparrow\downarrow}(z,t)=\langle\downarrow$$|_a\langle
  0|\hat{\Sigma}_{ge_-}(z)|\psi(t)\rangle$, and
  $S_{\uparrow\downarrow}(z,t)=\langle\downarrow$$|_a\langle
  0|\hat{\Sigma}_{gr}(z)|\psi(t)\rangle$}\BibitemShut {NoStop}%
\bibitem [{not()}]{note}%
  \BibitemOpen
  \href@noop {} {}\bibinfo {note} {Here we assume that the scattering
  coefficient $T$ is real. For a more general case where $T$ is a complex
  number, $\tilde{\rho}(x,y)$ is given by Eq.~(\ref{eq:eq7}) in regions $x\geq
  y$, and is modified by replacing $T$ with $T^*$ for $x\leq y$}\BibitemShut
  {NoStop}%
\end{thebibliography}

\end{document}


\preprint{APS/123-QED}

\title{Supplementary Material for ``Atom-Photon Spin-Exchange Collisions Mediated by Rydberg Dressing''}
\author{Fan Yang}
\affiliation{State Key Laboratory of Low Dimensional Quantum Physics, Department of Physics, Tsinghua University, Beijing 100084, China}
\author{Yong-Chun Liu}
\affiliation{State Key Laboratory of Low Dimensional Quantum Physics, Department of Physics, Tsinghua University, Beijing 100084, China}
\author{Li You}
\affiliation{State Key Laboratory of Low Dimensional Quantum Physics, Department of Physics, Tsinghua University, Beijing 100084, China}
\affiliation{Frontier Science Center for Quantum Information, Beijing 100084, China}
\affiliation{Beijing Academy of Quantum Information Sciences, Beijing 100193, China}
\maketitle
\onecolumngrid

This supplementary provides technical details of the main text, including: (i) derivation of the atom-photon interaction Hamiltonian (Sec.~\ref{sec:sec1}); (ii) analysis of the single-photon (Sec.~\ref{sec:sec2}) and multi-photon scattering dynamics (Sec.~\ref{sec:sec3}); (iii) considerations for experimental realizations (Sec.~\ref{sec:sec4}).

\section{derivation of the atom-photon interaction Hamiltonian}\label{sec:sec1}
In this section, we derive the effective Hamiltonian that governs the atom-photon interacting dynamics. We first consider the interaction between the control atom and ensemble atoms. For the level structure shown in Fig.~1(b) of the main text, atoms in states $|r\rangle$ and $|s\rangle$ are governed by the Hamiltonian
\begin{equation}
\hat{H} = \sum_n\left[E_r\hat{\sigma}_{rr}^n+E_s\hat{\sigma}_{ss}^n
+\left(\Omega e^{-i\omega_{L}t+i\mathbf{k}\cdot\mathbf{r}_n}\hat{\sigma}_{rs}^n+\mathrm{H.c.}\right)\right]
+\sum_{m<n}V_{mn}\hat{\sigma}_{rr}^m\hat{\sigma}_{rr}^n,
\label{eq:eqs1}
\end{equation}
where $V_{mn}=C_6/|\mathbf{r}_m-\mathbf{r}_n|^6$ ($C_6>0$) is the van der Walls (vdW) interaction between atoms in Rydberg state $|r\rangle$. In the large detuning regime $\Omega\ll\Delta=E_{r}-E_{s}-\omega_L$, the dynamics of a single Rydberg excitation can be described by
\begin{equation}
\label{eq:eqs2}
\hat{H}_\mathrm{eff} =\sum_{n}\left[\left(E_r+\delta\right)\hat{\sigma}_{rr}^n+\left(E_s-\delta\right)\hat{\sigma}_{ss}^n\right] +
\sum_{m\neq n} U_{mn}\left[\hat{\sigma}_{rr}^m\hat{\sigma}_{ss}^n + e^{i\mathbf{k}\cdot(\mathbf{r}_m-\mathbf{r}_n)}\hat{\sigma}_{rs}^m\hat{\sigma}_{sr}^n\right],
\end{equation}
where $\delta={\Omega^2}/{\Delta}$ is the linear light shift, and $U_{mn} = {\Omega^2V_{mn}}/[\Delta(\Delta+V_{mn})]$ denotes the Rydberg dressing indcued interaction \cite{yang2019quantum}. In the low-photon-density regime, the number of ensemble atoms excited to state $|r\rangle,|s\rangle$ is much smaller than one, so we can neglect the interaction between ensemble atom themselves and decompose Eq.~(\ref{eq:eqs2}) into $\hat{H}_\mathrm{eff}=\hat{H}_\mathrm{c}+\hat{H}_\mathrm{e}+\hat{H}_\mathrm{ce}$, where $\hat{H}_\mathrm{c}$, $\hat{H}_\mathrm{e}$, and $\hat{H}_\mathrm{ce}$ describe the Hamiltonian of the control atom, the ensemble atoms, and their mutual interactions, respectively. Dropping the superscript of the control atom and setting its location as the origin of the coordinate, we then obtain $\hat{H}_\mathrm{c} =\left(E_r+\delta\right)\hat{\sigma}_{rr}
+\left(E_s-\delta\right)\hat{\sigma}_{ss}$, $\hat{H}_\mathrm{e}=\sum_{n}\left[\left(E_r+\delta\right)\hat{\sigma}_{rr}^n
+\left(E_s-\delta\right)\hat{\sigma}_{ss}^n\right]$, and $\hat{H}_\mathrm{ce} = \sum_{n} U(\mathbf{r}_n)\left[\hat{\sigma}_{rr}\hat{\sigma}_{ss}^n + \hat{\sigma}_{ss}\hat{\sigma}_{rr}^n + ( e^{-i\mathbf{k}\cdot\mathbf{r}_n}\hat{\sigma}_{rs}\hat{\sigma}_{sr}^n+\mathrm{H.c.})\right]
$.

To describe collective excitations of ensemble atoms into state $|r\rangle,|s\rangle$, we introduce the collective spin operator $\hat{\Sigma}_{\mu\nu}(\mathbf{r}_n)=\sum_{n\in \Delta(\mathbf{r}_n)}\hat{\sigma}^n_{\mu\nu}/\sqrt{\rho(\mathbf{r}_n)}\Delta(\mathbf{r}_n)$ for atoms located within a small volume $\Delta(\mathbf{r}_n)$ around coarse grained $\mathbf{r}_n$ with a density $\rho(\mathbf{r}_n)$. Then, the state of ensemble atoms in $\mathbf{r}_n$ piece can be expressed as
\begin{equation}
|s^jr^k\rangle_{\mathbf{r}_n}=\sqrt{[\Delta(\mathbf{r}_n)]^{(j+k)}(N-j-k)!\hspace{1pt}N^{(j+k)}/(N!\hspace{1pt}j!\hspace{1pt}k!)}
\left[\hat{\Sigma}_{gs}^\dagger(\mathbf{r}_n)\right]^j\left[\hat{\Sigma}_{gr}^\dagger(\mathbf{r}_n)\right]^k|G\rangle,
\end{equation}
with $N = \rho(\mathbf{r}_n)\Delta(\mathbf{r}_n)$ and $|G\rangle=|g_1g_2\cdots g_{N}\rangle$. In the linear regime where the input photon number per $\mathbf{r}_n$ cell is much smaller than the corresponding atom $N$, or most atoms reside in the ground state $|g\rangle$, i.e., $j,k\ll N$. This makes collective excitations behave as bosonic quasi particles with $[\hat{\Sigma}_{g\mu}(\mathbf{r}_m),\hat{\Sigma}_{g\nu}^\dagger(\mathbf{r}_n)]
\approx\delta_{mn}\delta_{\mu\nu}/\Delta(\mathbf{r}_n)$, $[\hat{\Sigma}_{g\mu}^\dagger(\mathbf{r}_m),\hat{\Sigma}_{g\nu}^\dagger(\mathbf{r}_n)]
=[\hat{\Sigma}_{g\mu}(\mathbf{r}_m),\hat{\Sigma}_{g\nu}(\mathbf{r}_n)]
=0$, and $\hspace{1pt}\hat{\Sigma}_{\mu\nu}(\mathbf{r}_n)\approx
\hat{\Sigma}_{g\mu}^\dagger(\mathbf{r}_n)\hat{\Sigma}_{g\nu}(\mathbf{r}_n)/\sqrt{\rho(\mathbf{r}_n)} $ ($\mu,\nu=r,s$). The Hamiltonian involving ensemble atoms can then be expressed in terms of these collective bosonic operators as
\begin{align}
\hat{H}_\mathrm{e}&=\sum_{\mathbf{r}_n}\Delta(\mathbf{r}_n)\left[\left(E_r+\delta\right)\hat{\Sigma}_{gr}^\dagger(\mathbf{r}_n)\hat{\Sigma}_{gr}(\mathbf{r}_n)
+\left(E_s-\delta\right)\hat{\Sigma}_{gs}^\dagger(\mathbf{r}_n)\hat{\Sigma}_{gs}(\mathbf{r}_n)\right], \\
\hat{H}_\mathrm{ce} &=\sum_{\mathbf{r}_n} \Delta(\mathbf{r}_n) U(\mathbf{r}_n)\left[\hat{\sigma}_{rr}\hat{\Sigma}_{gs}^\dagger(\mathbf{r}_n)\hat{\Sigma}_{gs}(\mathbf{r}_n) +\hat{\sigma}_{ss}\hat{\Sigma}_{gr}^\dagger(\mathbf{r}_n)\hat{\Sigma}_{gr}(\mathbf{r}_n)+ \left(e^{-i\mathbf{k}\cdot\mathbf{r}_n}\hat{\sigma}_{rs}\hat{\Sigma}_{gs}^\dagger(\mathbf{r}_n)\hat{\Sigma}_{gr}(\mathbf{r}_n)+\mathrm{H.c.}\right)\right].
\end{align}
In the continuous limit $\Delta(\mathbf{r}_n)\rightarrow0$, the collective spin operators are replaced with bosonic spin-wave field operators satisfying $[\hat{\Sigma}_{g\mu}(\mathbf{r}),\hat{\Sigma}_{g\nu}^\dagger(\mathbf{r}^\prime)]
\approx\delta(\mathbf{r}-\mathbf{r}^\prime)\delta_{\mu\nu}$, and the summation $\sum_{\mathbf{r}_n} \Delta(\mathbf{r}_n)$
is replaced by the integral $\int d\mathbf{r}$.

With similar procedures, we obtain the Hamiltonian $\hat{H}_\mathrm{EIT}$ describing the double EIT couplings
\begin{align}
\hat{H}_\mathrm{EIT}&=\int d\mathbf{r}\left[g_pe^{i(\mathbf{k} _{\downarrow}\cdot \mathbf{r}-\nu_{\downarrow}t)}\hat{\mathcal{E}}_\downarrow(\mathbf{r})\hat{\Sigma}_{ge_+}^\dagger(\mathbf{r})+\Omega_\downarrow e^{i(\mathbf{q}_\downarrow\cdot\mathbf{r}-\omega_\downarrow t)}\hat{\Sigma}_{ge_+}^\dagger(\mathbf{r})\hat{\Sigma}_{gs}(\mathbf{r})
+\mathrm{H.c.}\right]\nonumber\\
&+\int d\mathbf{r}\left[g_pe^{i(\mathbf{k}_{\uparrow}\cdot \mathbf{r}-\nu_{\uparrow}t)}\hat{\mathcal{E}}_\uparrow(\mathbf{r})\hat{\Sigma}_{ge_-}^\dagger(\mathbf{r})+\Omega_\uparrow e^{i(\mathbf{q}_\uparrow\cdot\mathbf{r}-\omega_\uparrow t)}\hat{\Sigma}_{gr}^\dagger(\mathbf{r})\hat{\Sigma}_{ge_-}(\mathbf{r})
+\mathrm{H.c.}\right],
\end{align}
where $\hat{\mathcal{E}}_{\mu}(\mathbf{r})$ denotes the slowly varying operator for the quantized photonic field, $g_p$ is the collective atom-photon coupling constant for a uniform atomic density, and $\Omega_{\mu}$ denotes the Rabi frequency of the classical control field. We then transform the dynamics into the slowly varying and rotating frame according to $\hat{U}=\hat{U}_\downarrow\hat{U}_\uparrow\hat{U}_\mathrm{c}$ with
\begin{align}
\hat{U}_\downarrow=\exp\left\{i\int d\mathbf{r}
\left(\mathbf{k}_{\downarrow}\cdot\mathbf{r}-\nu_{\downarrow}t\right)
\hat{\Sigma}_{ge_+}^\dagger(\mathbf{r})\hat{\Sigma}_{ge_+}(\mathbf{r})+i\int d\mathbf{r} \left[(\mathbf{k}_{\downarrow}-\mathbf{q}_{\downarrow})\cdot\mathbf{r}-(\nu_{\downarrow}-\omega_{\downarrow})t\right]
\hat{\Sigma}_{gs}^\dagger(\mathbf{r})\hat{\Sigma}_{gs}(\mathbf{r})
\right\},\\
\hat{U}_\uparrow=\exp\left\{i\int d\mathbf{r}
\left(\mathbf{k}_{\uparrow}\cdot\mathbf{r}-\nu_{\uparrow}t\right)
\hat{\Sigma}_{ge_-}^\dagger(\mathbf{r})\hat{\Sigma}_{ge_-}(\mathbf{r})+i\int d\mathbf{r} \left[(\mathbf{k}_{\uparrow}+\mathbf{q}_{\uparrow})\cdot\mathbf{r}-(\nu_{\uparrow}+\omega_{\uparrow})t\right]
\hat{\Sigma}_{gr}^\dagger(\mathbf{r})\hat{\Sigma}_{gr}(\mathbf{r})
\right\},
\end{align}
and $\hat{U}_\mathrm{c}=\exp[-i(\nu_\uparrow+\omega_\uparrow)t\hat{\sigma}_{rr}
-i(\nu_\downarrow-\omega_\downarrow)t\hat{\sigma}_{ss}]$. For on-resonant EIT coupings $\nu_\downarrow=E_{e_+}$, $\nu_\uparrow=E_{e_-}$, $\delta_\downarrow=E_s-(\nu_\downarrow-\omega_\downarrow)=\delta$, and $\delta_\uparrow=E_r-(\nu_\uparrow+\omega_\uparrow)=-\delta$, we finally arrive at the total Hamiltonian $\hat{H}=\hat{H}_\mathrm{ph}+\hat{H}_\mathrm{EIT}+\hat{H}_\mathrm{ce}$ with
\begin{align}
\hat{H}_\mathrm{ph}&=\int{d\mathbf{r}}\left[-ic\hat{\mathcal{E}}_\downarrow^\dagger(\mathbf{r})\partial_z\hat{\mathcal{E}}_\downarrow(\mathbf{r})
-ic\hat{\mathcal{E}}_\uparrow^\dagger(\mathbf{r})\partial_z\hat{\mathcal{E}}_\uparrow(\mathbf{r})\right],
\label{eq:eqs9}\\
\hat{H}_\mathrm{EIT}&=\int d\mathbf{r}\left[g_p\hat{\mathcal{E}}_\downarrow(\mathbf{r})\hat{\Sigma}_{ge_+}^\dagger(\mathbf{r})+\Omega_\downarrow \hat{\Sigma}_{ge_+}^\dagger(\mathbf{r})\hat{\Sigma}_{gs}(\mathbf{r})+
g_p\hat{\mathcal{E}}_\uparrow(\mathbf{r})\hat{\Sigma}_{ge_-}^\dagger(\mathbf{r})+\Omega_\uparrow \hat{\Sigma}_{gr}^\dagger(\mathbf{r})\hat{\Sigma}_{ge_-}(\mathbf{r})
+\mathrm{H.c.}\right],\label{eq:eqs10}\\
\hat{H}_\mathrm{ce} &=\int d\mathbf{r}\hspace{1pt} U(\mathbf{r})\left[\hat{\sigma}_{rr}\hat{\Sigma}_{gs}^\dagger(\mathbf{r})\hat{\Sigma}_{gs}(\mathbf{r}) +\hat{\sigma}_{ss}\hat{\Sigma}_{gr}^\dagger(\mathbf{r})\hat{\Sigma}_{gr}(\mathbf{r})+ \left(\hat{\sigma}_{rs}\hat{\Sigma}_{gs}^\dagger(\mathbf{r})\hat{\Sigma}_{gr}(\mathbf{r})+\mathrm{H.c.}\right)\right],\label{eq:eqs11}
\end{align}
where we consider the two quantized photonic fields $\hat{\mathcal{E}}_\downarrow$ and $\hat{\mathcal{E}}_\uparrow$ copropagating along $z$-direction, neglect the diffraction in $x,y$ directions, and assume the momentum-matching condition $\mathbf{k}_\uparrow+\mathbf{q}_\uparrow=\mathbf{k}_\downarrow-\mathbf{q}_\downarrow+\mathbf{k}$ is satisfied. The following analysis of atom-photon collisional dynamics is based on Eqs.~(\ref{eq:eqs9})-(\ref{eq:eqs11}).

\section{Analysis of single-photon scattering}\label{sec:sec2}
We first consider the scattering dynamics of a single photon. The quantum state for the single control atom and a single photon (or a single spin-wave excitation) can be written as
\begin{align}
|\psi(t)\rangle&=\int{d\mathbf{r}}\left[E_{\downarrow\uparrow}(\mathbf{r},t)\hat{\mathcal{E}}_{\downarrow}^\dagger(\mathbf{r})|0\rangle
+P_{\downarrow\uparrow}(\mathbf{r},t)\hat{\Sigma}_{ge_+}^\dagger(\mathbf{r})|0\rangle
+S_{\downarrow\uparrow}(\mathbf{r},t)\hat{\Sigma}_{gs}^\dagger(\mathbf{r})|0\rangle\right]\otimes|\hspace{-2.5pt}\uparrow\rangle_{a}\nonumber\\
&+\int{d\mathbf{r}}\left[E_{\uparrow\downarrow}(\mathbf{r},t)\hat{\mathcal{E}}_{\uparrow}^\dagger(\mathbf{r})|0\rangle
+P_{\uparrow\downarrow}(\mathbf{r},t)\hat{\Sigma}_{ge_-}^\dagger(\mathbf{r})|0\rangle
+S_{\uparrow\downarrow}(\mathbf{r},t)\hat{\Sigma}_{gr}^\dagger(\mathbf{r})|0\rangle\right]\otimes|\hspace{-2.5pt}\downarrow\rangle_{a},\label{eq:eqs17}
\end{align}
where $|\hspace{-2.5pt}\uparrow\rangle_{a}=|r\rangle$ and $|\hspace{-2.5pt}\downarrow\rangle_{a}=|s\rangle$ denote the state of the control atom. The evolution of the wavefunctions in Schr\"{o}dinger picture can be determined by studying the Heisenberg equations for the field operator
\begin{align}
\partial_t\hat{\mathcal{E}}_{\downarrow}(\mathbf{r})&=-c\partial_z\hat{\mathcal{E}}_{\downarrow}(\mathbf{r})-ig_p\hat{\Sigma}_{ge_+}(\mathbf{r}),\label{eq:eqs18}\\
\partial_t\hat{\Sigma}_{ge_+}(\mathbf{r})&=-\gamma\hat{\Sigma}_{ge_+}(\mathbf{r})-ig_p\hat{\mathcal{E}}_{\downarrow}(\mathbf{r})-i\Omega_\downarrow\hat{\Sigma}_{gs}(\mathbf{r})+\hat{F}_{e_+}(\mathbf{r}),\label{eq:eqs19}\\
\partial_t\hat{\Sigma}_{gs}(\mathbf{r})&=-i\Omega_\downarrow\hat{\Sigma}_{ge_+}(\mathbf{r})-iU(\mathbf{r})\hat{\sigma}_{rr}\hat{\Sigma}_{gs}(\mathbf{r})-iU(\mathbf{r})\hat{\sigma}_{rs}\hat{\Sigma}_{gr}(\mathbf{r}),\label{eq:eqs20}\\
\partial_t\hat{\mathcal{E}}_{\uparrow}(\mathbf{r})&=- c\partial_z\hat{\mathcal{E}}_{\uparrow}(\mathbf{r})-ig_p\hat{\Sigma}_{ge_-}(\mathbf{r}),\label{eq:eqs21}\\
\partial_t\hat{\Sigma}_{ge_-}(\mathbf{r})&=-\gamma\hat{\Sigma}_{ge_-}(\mathbf{r})-ig_p\hat{\mathcal{E}}_{\uparrow}(\mathbf{r})-i\Omega_\uparrow\hat{\Sigma}_{gr}(\mathbf{r})+\hat{F}_{e_-}(\mathbf{r}),\label{eq:eqs22}\\
\partial_t\hat{\Sigma}_{gr}(\mathbf{r})&=-i\Omega_\uparrow\hat{\Sigma}_{ge_-}(\mathbf{r})-iU(\mathbf{r})\hat{\sigma}_{ss}\hat{\Sigma}_{gr}(\mathbf{r})-iU(\mathbf{r})\hat{\sigma}_{sr}\hat{\Sigma}_{gs}(\mathbf{r}),\label{eq:eqs23}
\end{align}
where $2\gamma$ is the linewidth of the intermediate state $|e_\pm\rangle$ and $\hat{F}_{e_\pm}(\mathbf{r})$ describes the associated Langevin noise (which does not affect the calculation of the single-excitation wavefunction). With Eqs.~(\ref{eq:eqs18})-(\ref{eq:eqs23}), we then obtain the equations of motion for the wavefunctions in Eq.~(\ref{eq:eqs17}) [Eq.~(2) in the main text].

The above derivation holds for scatterings in 3D, while in the main text we focus on the scattering in 1D case. For 3D case, the scattering coefficients $T(\omega)$ and $R(\omega)$ are defined as
\begin{equation}
T(\omega)=\frac{\int d\mathbf{r}_\perp{E}_{\downarrow\uparrow}^*(\mathbf{r}_\perp,z=0,\omega){E}_{\downarrow\uparrow}(\mathbf{r}_\perp,z=L,\omega)}
{\int d\mathbf{r}_\perp|{E}_{\downarrow\uparrow}(\mathbf{r}_\perp,z=0,\omega)|^2},\quad
R(\omega)=\frac{\int d\mathbf{r}_\perp{E}_{\downarrow\uparrow}^*(\mathbf{r}_\perp,z=0,\omega){E}_{\uparrow\downarrow}(\mathbf{r}_\perp,z=L,\omega)}
{\int d\mathbf{r}_\perp|{E}_{\downarrow\uparrow}(\mathbf{r}_\perp,z=0,\omega)|^2},
\end{equation}
where $\mathbf{r}_\perp=\{x,y\}$ denotes transverse coordinates, and ${E}_{\downarrow\uparrow}(\mathbf{r}_\perp,z=0,\omega)=\exp[-|\mathbf{r}_\perp|^2/2w-(\omega\Delta t)^2/2]\sqrt{4\Delta t/c\sqrt{\pi}}/w$ denotes the (frequency-domain) wavefunction of a Gaussian input beam with a waist $w$ and a temporal width $\Delta t$.

\subsection{Comparison with the scheme using off-diagonal vdW interactions}
\begin{figure}[b]
	\centering
	\includegraphics[width=0.85\linewidth]{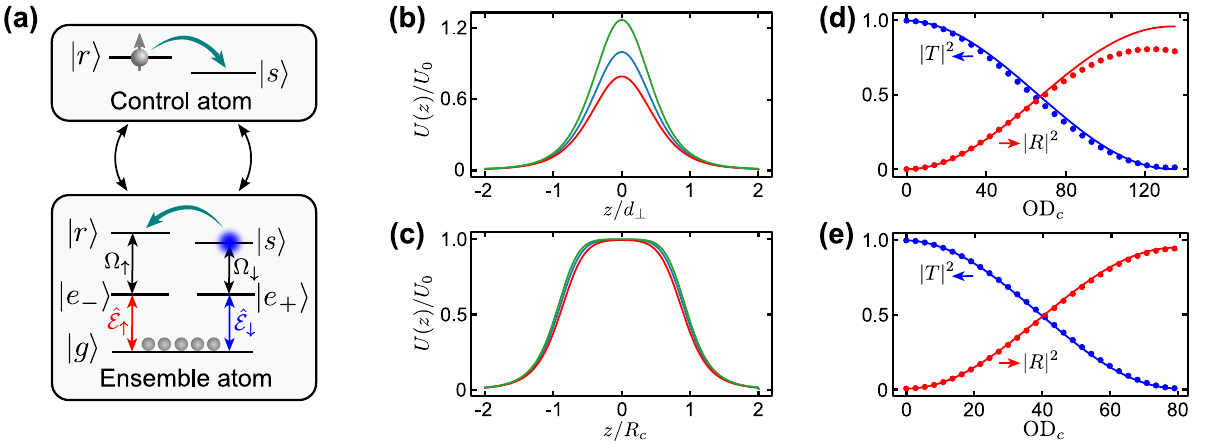}
	\caption{(a) Level diagram of the control atom and ensemble atoms for the scheme using off-diagonal vdW interactions. (b) and (c) show the effective potential $U(z)$ for the scheme using off-diagonal vdW interactions and the dressing scheme, respectively, with $\Delta r_\perp=1~\mu\mathrm{m}$ (red lines), $0~\mu\mathrm{m}$ (blue lines), $-1~\mu\mathrm{m}$ (green lines). (d) and (e) show the scattering coefficients versus the effective optical depth $\mathrm{OD}_c$. The dots and the solid lines correspond to the results for a Gaussian beam with a waist $w$ and the $1$D model. The parameters used are the same as in Fig.~2(c) of the main text, which gives $d_\perp=25~\mu\mathrm{m}$ for the scheme using off-diagonal vdW interactions, and $d_\perp=4~\mu\mathrm{m}$, $R_c=12~\mu\mathrm{m}$ for the dressing scheme.}
	\label{fig:figs1}
\end{figure}

In addition to Rydberg dressing, the spin-exchange interaction can also be induced by exploiting off-diagonal vdW interaction between two Rydberg states. As shown in Fig.~\ref{fig:figs1}(a), state $|s\rangle$ now denotes another Rydberg state (e.g., $|s\rangle = |n^\prime S_{1/2}, J =1/2, m_J = -1/2\rangle$ of $^{87}\mathrm{Rb}$ atom). In this configuration, the Hamiltonian describing the interaction between the control atom and ensemble atoms is $\hat{H}_\mathrm{ce} = \sum_{n} U(\mathbf{r}_n)\left[\hat{\sigma}_{rr}\hat{\sigma}_{ss}^n + \hat{\sigma}_{ss}\hat{\sigma}_{rr}^n + \lambda(\hat{\sigma}_{rs}\hat{\sigma}_{sr}^n+\mathrm{H.c.})\right]$ \cite{glaetzle2017quantum}. Near a F{\"o}rster resonance, $\lambda\approx1$, which gives a total Hamiltonian of the same form as Eqs.~(\ref{eq:eqs9})-(\ref{eq:eqs11}). In this case, the interaction potential is replaced by $U(\mathbf{r})=U_0/[\sqrt{z^2+r_\perp^2}/d_\perp]^6$, where $d_\perp$ denotes the transverse separation between the control atom and the center of the input beam, and $U_0=C_6/d_\perp^6$ denotes the maximum interaction strength. Correspondingly, the interaction-induced phase is approximately given by $\phi\approx(3\pi/8)\xi[1-i(21/32)\xi]\times\mathrm{OD}_c$, where the effective optical depth becomes $\mathrm{OD}_c=g_p^2d_\perp/\gamma c$, and $\xi=U_0/\gamma_\mathrm{EIT}$ also determines the scattering property. 

From the perspective of experimental realization, the scheme using off-diagonal vdW interaction is simpler than the dressing scheme discussed in the main text, as it only requires two control beams. Furthermore, since the direct off-diagonal vdW interaction is stronger than the dressing induced interaction, the separation $d_\perp$ can be made larger in this scheme, which reduces the crosstalk between the control atom and ensemble atoms [to be discussed in Sec.~\ref{sec:sec4}]. However, the performance of this scheme is found to be not as good as the dressing scheme. 

First, the off-diagonal vdW interaction is relatively sensitive to the variation of the transverse separation $\Delta r_\perp = r_\perp-d_\perp$ between the control atom and the input photon [Fig.~\ref{fig:figs1}(b)]. This will distort the transverse profile of the photonic wavefunction and result in a reduced fidelity compared with the prediction of the 1D model [Fig.~\ref{fig:figs1}(d)]. In contrast, the plateau of the dressed interaction potential makes it highly insensitive to the variation $\Delta r_\perp$ [Fig.~\ref{fig:figs1}(c)], and can keep the transverse mode profile of the transmitted photon unaltered [Fig.~\ref{fig:figs1}(e)].

Second, in this scheme, the strong interaction between Rydberg atoms in state $|s\rangle$ will induce a strong dissipative nonlinearity between input photons themselves for multi-photon scattering, which increases the nonlinear loss of the system. For the dressing scheme, such an unwanted direct interaction between input photons is suppressed by a factor of $(\Omega/\Delta)^2$ compared with the interaction strength between the control atom and the input photon.

\subsection{Long-distance entanglement with single-photon scatterings}
\label{subsection1}
As discussed above, the spin-exchange collision between the control atom in state $|\hspace{-2.5pt}\uparrow\rangle_a$ and a single input photon in mode $\hat{a}_\downarrow(t)=\int_{-\infty}^\infty dzh(z-ct)\hat{\mathcal{E}}_\downarrow^\dagger(z)$ results in an entangled output state
\begin{equation}
|\psi_a(t)\rangle=\frac{1}{\sqrt{p}}\left(T\hat{a}^\dagger_\downarrow(t)|0\rangle\otimes|\hspace{-2pt}\uparrow\rangle_a+R\hat{a}^\dagger_\uparrow(t)|0\rangle\otimes|\hspace{-2.5pt}\downarrow\rangle_a\right), \end{equation}
conditioned on the survival of the input photon, with a success probability $p=|T|^2+|R|^2$. We now discuss the implementation of the quantum repeater protocol using our system.

\begin{figure}
\centering
\includegraphics[width=0.75\linewidth]{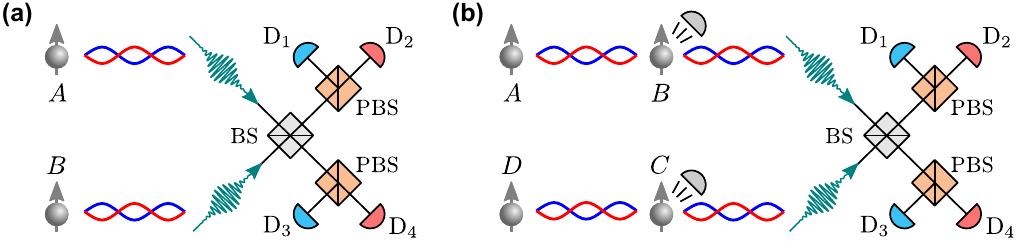}
\caption{(a) Schematic for building up the elementary entanglement between node $A$ and node $B$. (b) Schematic for implementing the entanglement connection between node $A$ and node $D$.}
\label{fig:figs2}
\end{figure}

First, we describe how to create elementary entanglement. As shown in Fig.~\ref{fig:figs2}(a), two independent spin-exchange collisions at node $A$ and node $B$ will produce a product state $|\psi_a\rangle\otimes|\psi_b\rangle$. The output photons from node $A$ and node $B$ are then combined at a beam splitter (BS), which transforms optical modes as $\hat{a}_\mu\rightarrow(\hat{a}_\mu+e^{i\varphi}\hat{b}_\mu)/\sqrt{2}$ and
$\hat{b}_\mu\rightarrow(\hat{b}_\mu-e^{-i\varphi}\hat{a}_\mu)/\sqrt{2}$ with $\mu=\uparrow,\downarrow$, and yields an output state
\begin{equation}
|\psi_{ab}\rangle=\frac{TR}{\sqrt{2}p}\left(
e^{i\varphi}\hat{b}^\dagger_\uparrow\hat{b}^\dagger_\downarrow|0\rangle\otimes|\Phi_+\rangle
-e^{-i\varphi}\hat{a}^\dagger_\uparrow\hat{a}^\dagger_\downarrow|0\rangle\otimes|\Phi_+\rangle
-\hat{a}^\dagger_\uparrow\hat{b}^\dagger_\downarrow|0\rangle\otimes|\Phi_-\rangle
+\hat{a}^\dagger_\downarrow\hat{b}^\dagger_\uparrow|0\rangle\otimes|\Phi_-\rangle
\right)+|\psi_\mathrm{dis}\rangle,
\end{equation}
where $|\Phi_\pm\rangle=(|\hspace{-2.5pt}\uparrow\rangle_a|\hspace{-2pt}\downarrow\rangle_b
\pm|\hspace{-2.5pt}\downarrow\rangle_a|\hspace{-2pt}\uparrow\rangle_b)/\sqrt{2}$ denotes the Bell states of atoms, $|\psi_\mathrm{dis}\rangle$ represents the state where output photons carry the same spins and will be discarded after postselection. With additional polarization beam splitters (PBS), we can detect photons in different modes, e.g., $\mathrm{D}_1,\mathrm{D}_2,\mathrm{D}_3,\mathrm{D}_4$ for registering photons in mode $\hat{b}_\downarrow,\hat{b}_\uparrow,\hat{a}_\downarrow,\hat{a}_\uparrow$ respectively. The click of detectors $\{\mathrm{D}_1,\mathrm{D}_2\}$ or $\{\mathrm{D}_3,\mathrm{D}_4\}$ then heralds the generation of maximally entangled state $|\Phi_+\rangle$, while the click of $\{\mathrm{D}_1,\mathrm{D}_4\}$ or $\{\mathrm{D}_2,\mathrm{D}_3\}$ heralds the Bell state $|\Phi_-\rangle$.

Once the elementary entanglement is established, we can proceed to extend it to longer distance via entanglement connections. Suppose that we have two entangled pairs $|\Phi_+\rangle_{ab}$ and $|\Phi_+\rangle_{cd}$ distributed at $\{A,B\}$ and $\{C,D\}$, respectively. Then, we inject a spin-down single-photon $\hat{b}^\dagger_\downarrow|0\rangle$ interacting with the atom at node $B$ and measure the spin state of the atom. Conditioned on the survival of the input photon and the measurement result $|$$\downarrow\rangle_b$, we can swap the atom-atom entanglement for the atom-photon entanglement, i.e., creating an entangled state
\begin{equation}
|\phi_{ab}\rangle=\frac{1}{\sqrt{2(1+|R|^2)}}\left(\hat{b}^\dagger_\downarrow|0\rangle\otimes|\hspace{-2.5pt}\uparrow\rangle_a
+R\hat{b}^\dagger_\uparrow|0\rangle\otimes|\hspace{-2.5pt}\downarrow\rangle_a\right)
\end{equation}
with a success probability $(1+|R|^2)/2$. Similarly, we can exchange the state of the atom at node $C$ with a single photon probabilistically, and transform the atom entangled pair $|\Phi_+\rangle_{cd}$ into the atom-photon entangled state $|\phi_{dc}\rangle$. Then, by using the same linear optics setup described previously [see Fig.~\ref{fig:figs2}(b)], we can herald that atoms at nodes $A$ and $D$ are entangled in Bell basis $|\Phi_+\rangle_{cd}$ or $|\Phi_-\rangle_{cd}$, depending on the outcomes of the photon detections.

There are several advantages of our scheme compared with the popular DLCZ protocol \cite{duan2001long}. First, the entanglement here is encoded in atomic spin basis rather than in the Fock space, such that photon-number resolved detectors are not required \cite{sangouard2011quantum}. Second, the entangled state produced here is independent of the phase $\varphi$. Such an interferometric phase is always unstable over long communication time scales, and can severely limit the performance of quantum repeaters based on single photon detections \cite{chen2007fault}. Further, the quantum information at each node is encoded in the internal state of a single atom, which is much easier to manipulate and to detect compared with an atomic ensemble, and does not require the quantum storage and retrieval.

The above scheme remains workable in the presence of photon loss and imperfect detection efficiency. For a more realistic consideration, dark counts of photon detectors and decoherence of atomic state will reduce the fidelity of the final Bell state, which can be further optimized by using error-correction procedures \cite{sangouard2011quantum}.

\section{Analysis of Multi-photon scattering}\label{sec:sec3}
We consider the multi-photon scattering in the coherent regime ($\xi\ll1$ and $|T|^2+|R|^2\approx1$). As verified in the main text, the dynamics in this case can be described by a Hermitian Hamiltonian
\begin{equation}
\label{eq:eqs26}
\hat{H}=\sum_{\mu=\downarrow,\uparrow}-iv_{\mu}\int dz\hat{\Psi}^\dagger_{\mu}(z)\partial_z\hat{\Psi}_{\mu}(z)+\int{dz}U(z)\left[\hat{\sigma}_{\uparrow\uparrow}\hat{\Psi}_{\downarrow}^\dagger(z)\hat{\Psi}_{\downarrow}(z) +\hat{\sigma}_{\downarrow\downarrow}\hat{\Psi}_{\uparrow}^\dagger(z)\hat{\Psi}_{\uparrow}(z)+ \left(\hat{\sigma}_{\uparrow\downarrow}\hat{\Psi}_{\downarrow}^\dagger(z)\hat{\Psi}_{\uparrow}(z)+\mathrm{H.c.}\right)\right].
\end{equation}
The symmetry of this Hamiltonian and the low-photon-density assumption allows us to obtain an analytical form of the output state $|\psi\rangle$, as given by Eq.~(6) in the main text. To analyze such a many-body entangled state, we trace out remaining photons in mode $\hat{\mathcal{E}}_\downarrow$ to obtain a reduced density matrix $\hat{\rho}=\int dxdy\rho(x,y)\hat{\mathcal{E}}_{\uparrow}^\dagger(x)|0\rangle\langle 0|\hat{\mathcal{E}}_{\uparrow}(y)$ for the single photon in mode $\hat{\mathcal{E}}_\uparrow$, where the matrix element $\rho(x,y)$ is determined by
\begin{equation}
\rho(x,y)=\sum_{m=0}^{\infty}\frac{1}{m!}\int dz_{1}dz_{2}\cdots{dz_m}\langle\downarrow\hspace{-2.5pt}|_a\langle0|\hat{\mathcal{E}}_\downarrow(z_1)\hat{\mathcal{E}}_\downarrow(z_2)\cdots
\hat{\mathcal{E}}_\downarrow(z_m)\hat{\mathcal{E}}_{\uparrow}(x)|\psi\rangle\langle\psi|\hat{\mathcal{E}}_{\uparrow}^\dagger(y)\hat{\mathcal{E}}_\downarrow^\dagger(z_1)\hat{\mathcal{E}}_\downarrow^\dagger(z_2)\cdots
\hat{\mathcal{E}}_\downarrow^\dagger(z_m)|0\rangle|\hspace{-2.5pt}\downarrow\rangle_a,
\end{equation}
The density matrix element $\rho(x,y)$ for the $n$-photon Fock input state is given in the main text. For a coherent input state $e^{-|\alpha|^2/2}\sum_n(\alpha^n/\sqrt{n!})|n\rangle$, we find $\rho(x,y)=\tilde{\rho}(t-\tau-x,t-\tau-y)$ with
\begin{equation}
\tilde{\rho}(x,y)=|\alpha R|^2h(x)h(y) \exp{\left[-|\alpha|^2 |R|^2\int_{-\infty}^{y}dzh^2(z)-|\alpha|^2(1-T)\int_{y}^{x}dzh^2(z)\right]}.
\end{equation}
when $x>y$ and $\tilde{\rho}(x,y)=\tilde{\rho}^*(y,x)$ for $x\leq y$. The efficiency and the purity for the extracted single photon are respectively given by $\eta=\mathrm{tr}[\hat{\rho}]=1-e^{-|\alpha|^2(1-|T|^2)}$ and
\begin{align}
\mathcal{P}=\frac{\mathrm{tr}[\hat{\rho}^2]}{\mathrm{tr}[\hat{\rho}]^2}=
\frac{|R|^4}{2\eta^2\left(\mathrm{Re}[T]-|T|^2\right)}\left[\frac{1-e^{-2|\alpha|^2(1-\mathrm{Re}[T])}}{1-\mathrm{Re}[T]}
-\frac{1-e^{-2|\alpha|^2(1-|T|^2)}}{1-|T|^2}\right].
\end{align}

To verify the analytical results, we perform numerical calculations for the $n=2$ Fock input state. The output state
\begin{equation}
|\psi(t)\rangle = \hspace{2pt}\frac{1}{2}\iint{dz_1}{dz_2}E_{\downarrow\downarrow\uparrow}(z_1,z_2,t)
\hat{\mathcal{E}}^\dagger_\downarrow(z_1)\hat{\mathcal{E}}^\dagger_\downarrow(z_2)|0\rangle|\hspace{-2.5pt}\uparrow\rangle_a
+\iint{dz_1}{dz_2}E_{\uparrow\downarrow\downarrow}(z_1,z_2,t)
\hat{\mathcal{E}}^\dagger_\uparrow(z_1)\hat{\mathcal{E}}^\dagger_\downarrow(z_2)|0\rangle|\hspace{-2.5pt}\downarrow\rangle_a
\end{equation}
can be obtained by solving equations of motion for DSP wavefunctions derived from Eq.~(\ref{eq:eqs26}),
\begin{align}
\partial_t\Psi_{\uparrow\downarrow\downarrow}(z_1,z_2,t)=&
-v_\uparrow\partial_{z_1}\Psi_{\uparrow\downarrow\downarrow}(z_1,z_2,t)
-iU(z_1)\Psi_{\uparrow\downarrow\downarrow}(z_1,z_2,t)
-iU(z_1)\Psi_{\downarrow\downarrow\uparrow}(z_1,z_2,t)-v_\downarrow\partial_{z_2}\Psi_{\uparrow\downarrow\downarrow}(z_1,z_2,t),\\
\partial_t\Psi_{\downarrow\uparrow\downarrow}(z_1,z_2,t)=&
-v_\downarrow\partial_{z_1}\Psi_{\downarrow\uparrow\downarrow}(z_1,z_2,t)-v_\uparrow\partial_{z_2}\Psi_{\downarrow\uparrow\downarrow}(z_1,z_2,t)
-iU(z_2)\Psi_{\downarrow\uparrow\downarrow}(z_1,z_2,t)
-iU(z_2)\Psi_{\downarrow\downarrow\uparrow}(z_1,z_2,t),\\
\partial_t\Psi_{\downarrow\downarrow\uparrow}(z_1,z_2,t)=&
-v_\downarrow\partial_{z_1}\Psi_{\downarrow\downarrow\uparrow}(z_1,z_2,t)
-iU(z_1)\Psi_{\downarrow\downarrow\uparrow}(z_1,z_2,t)
-iU(z_1)\Psi_{\uparrow\downarrow\downarrow}(z_1,z_2,t)
-v_\downarrow\partial_{z_2}\Psi_{\downarrow\downarrow\uparrow}(z_1,z_2,t)\nonumber\\
&-iU(z_2)\Psi_{\downarrow\downarrow\uparrow}(z_1,z_2,t)
-iU(z_2)\Psi_{\downarrow\uparrow\downarrow}(z_1,z_2,t),
\end{align}
together with the boundary condition ($\alpha,\lambda,\mu=\uparrow,\downarrow$)
\begin{align}
\Psi_{\alpha\lambda\mu}(z_1=0,z_2,t) &= E_{\alpha\lambda\mu}(z_1=0,z_2,t)\sqrt{c/v_\alpha},\quad\hspace{3pt} \Psi_{\alpha\lambda\mu}(z_1,z_2=0,t) = E_{\mu\nu\lambda}(z_1,z_2=0,t)\sqrt{c/v_\lambda},\\ E_{\alpha\lambda\mu}(z_1=L,z_2,t) &= \Psi_{\alpha\lambda\mu}(z_1=L,z_2,t)\sqrt{v_\alpha/c},\quad E_{\alpha\lambda\mu}(z_1,z_2=L,t)= \Psi_{\alpha\lambda\mu}(z_1,z_2=L,t)\sqrt{v_\lambda/c}.
\end{align}

\begin{figure}
	\centering
	\includegraphics[width=0.9\linewidth]{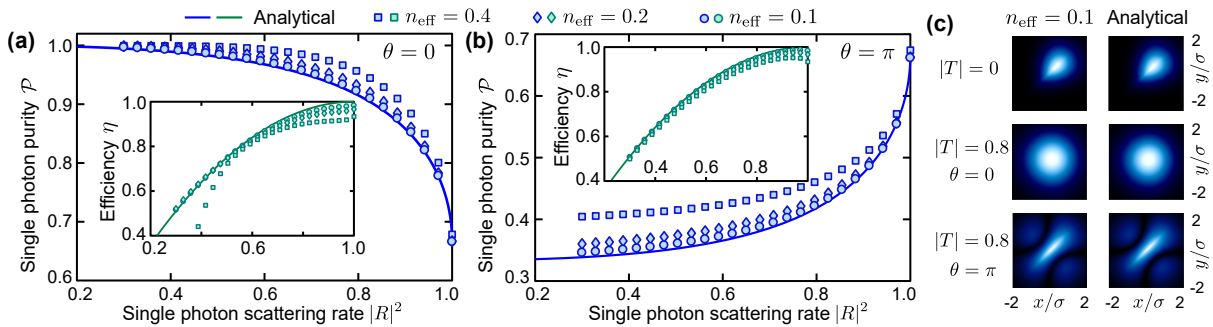}
	\caption{Purity and efficiency of the extracted photon for (a) $\theta=0$ (by taking $\phi=\pi/2$ and $\Omega_\downarrow>\Omega_\uparrow$) and (b) $\theta=\pi$ (by taking $\phi=\pi/2$ and $\Omega_\downarrow<\Omega_\uparrow$). (c) Normalized density matrix $|\rho(x,y)|/\eta$ for the indicated values of $T$. The left and the right columns show numerical calculations (with $n_\mathrm{eff}=0.1$) and analytical results, respectively.}
	\label{fig:figs3}
\end{figure}

The accuracy of the approximate analytical solution depends on the effective number of photons $n_\mathrm{eff}=nR_c/v_\downarrow\Delta t$ that can be simultaneously interacting with the control atom. As shown in Figs.~\ref{fig:figs3}(a) and \ref{fig:figs3}(b), as $n_\mathrm{eff}$ becomes smaller to validate the low-photon-density assumption $n_\mathrm{eff}\ll1$, the analytical predictions for purities and efficiencies show better agreement with the numerical results. In the main text, we take $n_\mathrm{eff}=0.1$, with which the density matrix $\rho(x,y)$ can be well described by the analytical solution [Fig.~\ref{fig:figs3}(c)]. For a large $n_\mathrm{eff}$, the analytical solution is not as accurate, but can still provide a qualitative prediction of the output state as verified by Fig.~\ref{fig:figs3}.

\section{Experimental considerations}\label{sec:sec4}
There are several aspects that need to be carefully considered for the experimental realization of our scheme. First, the validity of the 1D treatment requires the effective range $R_c$ of the potential to be larger than the waist and smaller than the Rayleigh range of the input beam, i.e., $w<R_c<\pi w^2/\lambda_0$. In Fig.~2(b) of the main text, we take $w=2~\mu\mathrm{m}$, $R_c=12~\mu\mathrm{m}$, and $\lambda_0=0.78~\mu\mathrm{m}$, which satisfies the above condition. It is worth pointing out that the deviation from this condition will not influence the mode profile of the output photon in the dissipative interacting regime at large $\mathrm{OD}_c$, as the entangled dark state does not depend on the detail of the interaction potential.

Second, the tail of the EIT pumping beams $\Omega_{\uparrow}$ and $\Omega_{\downarrow}$ can influence the control atom. To suppress such crosstalks, the waists of these pumping beams need to be smaller than the distance $d_\perp$ between the control atom and the center of the atomic ensemble. The crosstalks can be minimized by using different species of atoms for the control one and the ensemble one, where the dressing fields $\Omega$ for these two species of atoms also need to be different.

Third, the finite dressing parameter $(\Omega/\Delta)$ puts a limitation on the maximum photon number $n_\mathrm{max}\approx(\Delta/\Omega)^2$. For a typical dressing parameter $\Omega/\Delta=0.1$, the input photon number should be smaller than 100, otherwise there would on average be one photon whose spin is directly flipped without interacting with the control atom.

Finally, we discuss the necessary condition for neglecting the decay of the Rydberg state. First, the time duration for a photon inside the medium should be much smaller than the inverse decay rate $1/\gamma_s$ of the Rydberg collective excitation in the atomic ensemble. Second, the time duration for a photon completely passing through the medium needs to be much smaller than the inverse decay rate $1/\gamma_c$ of the Rydberg state for the control atom. These requirements lead to the conditions (i) $4\gamma_sR_c/v_\downarrow\ll1$; and (ii) $\gamma_c(\Delta t+4R_c/v_\downarrow)\ll1$, where $\Delta t$ denotes the time duration of the input pulse, and the length of the medium is taken to be $L=4R_c$. After the interaction, one can transfer the Rydberg excitation of the control atom to another long-lived ground state to preserve its coherence. For realistic decay rates $\gamma_s/2\pi=0.1~\mathrm{MHz}$ and $\gamma_c/2\pi=5~\mathrm{kHz}$ in current experimental setups, the parameters used in Fig.~2(c) of the main text give $4\gamma_sR_c/v_\downarrow\approx 0.66$ and $\gamma_c(\Delta t+4R_c/v_\downarrow)\approx 0.06$. Although condition (i) is not strictly satisfied for these experimental parameter choices, it only introduces certain extra photon loss, and one can make use of post-selection to explore the underlying physics of the system and some possible applications discussed in Sec.~\ref{subsection1}.

\bibliography{supplement}